\documentclass[prd,aps, preprintnumbers, showpacs, nofootinbib,superscriptaddress,notitlepage,twocolumn]{revtex4-1}
\usepackage[normalem]{ulem}
\usepackage{epsfig}
\usepackage{amsfonts}
\usepackage{amsmath}
\usepackage{slashed}
\usepackage{graphicx}
\usepackage{color}
\usepackage{mathtools}
\usepackage{simpler-wick}
\usepackage{simpler-wick}
\allowdisplaybreaks[4]

\newcommand{\beq}{\begin{eqnarray}}
\newcommand{\eeq}{\end{eqnarray}}

\begin{document}
	
	\preprint{JLAB-THY-20-3205}
	
	\title{$B$-meson Ioffe-time distribution amplitude at short distances }

	\author{Shuai Zhao}
	\email{szhao@odu.edu}
	\affiliation{Department of Physics, Old Dominion University, Norfolk, VA 23529, USA}
	\affiliation{Theory Center, Thomas Jefferson National Accelerator Facility, Newport News, VA 23606, USA}
	
		\author{Anatoly V. Radyushkin}
\email{radyush@jlab.org}
\affiliation{Department of Physics, Old Dominion University, Norfolk, VA 23529, USA}
\affiliation{Theory Center, Thomas Jefferson National Accelerator Facility, Newport News, VA 23606, USA}
	
	\begin{abstract}

		We propose  the  approach for a lattice investigation of   light-cone distribution amplitudes (LCDA) of heavy-light mesons, such as the $B$-meson, 
	using the formalism of parton pseudo-distributions.  A basic ingredient of the approach is  the   study 
	of  short-distance behavior of the $B$-meson Ioffe-time distribution amplitude (ITDA),   
	which is a generalization of the $B$-meson LCDA  in coordinate space. 
	We construct a reduced ITDA for the $B$-meson, and derive the matching relation between 
	the reduced ITDA and the LCDA.   The reduced ITDA is ultraviolet finite, which guarantees that   the continuum limit exists on the lattice. 

	\end{abstract}

	\date{\today}

	\maketitle
	
\section{Introduction}

The  $B$-meson  physics plays  a remarkable   role in particle physics,
	both  in  a detailed  examination of the Standard Model
	and  in  the search of new physics beyond the Standard Model.
One of the most important functions describing the structure of the $B$-meson 
is its light-cone distribution amplitude (LCDA)~\cite{Grozin:1996pq}. 
It is an inherent part  of  hard-collinear factorization theorems for many exclusive $B$ decay reactions~\mbox{\cite{Beneke:2000ry,Beneke:2001ev,Beneke:2001at,DescotesGenon:2002mw,Bauer:2002aj,Beneke:2003zv,Becher:2005fg},} where the amplitude is factorized into a convolution of the  hard scattering kernel and the $B$-meson LCDA.  It is also an essential element in 
 the  light-cone sum-rule
studies~\cite{Khodjamirian:2006st,Faller:2008tr,Gubernari:2018wyi,Wang:2015vgv,Wang:2017jow,Lu:2018cfc,Gao:2019lta}
of the $B$-meson decays.

 The perturbative structure   of the  $B$-meson LCDA 
 may be  studied in a model-independent way, e.g., using 
 the renormalization group equation~\cite{Lange:2003ff,Bell:2013tfa,Braun:2014owa,Braun:2019wyx} and  constraints 
 on the perturbative tail of the leading-twist LCDA 
 $\phi_B^+(\omega,\mu)$~\cite{Lee:2005gza,Feldmann:2014ika}.  
 On the other hand, the nonperturbative  aspects  of $B$-meson LCDA has been  mainly explored
 within models based on
 QCD sum rules~\cite{Braun:2003wx,Khodjamirian:2005ea}. 
 
 A  first-principle approach to study the nonperturbative 
 aspects of the $B$-meson LCDA  may be provided by lattice gauge simulations.
However, there was  
 not  much 
 work in this direction. 
 The difficulties arise from the fact that, in
 the
 heavy quark effective theory (HQET), the $B$-meson LCDA is defined through the matrix element of a
 nonlocal operator in which the heavy and light quarks are separated along the light-cone.
 
  Thus, it  cannot be  calculated directly on the Euclidean lattice. Moreover, unlike 
  in the case of the parton distribution functions of  the nucleon, it is impossible to access $B$-meson LCDA
by computing its moments, just  because the operator product expansion (OPE) does not exist
in this case~\cite{Braun:2003wx}. One might 
 propose 
to calculate  instead  the inverse moments of LCDA which  are more relevant to phenomenology. 
However, they are not related to matrix elements of local operators.

 The recent developments in the study of  parton distribution functions (PDFs) on the lattice  (e.g., quasi-PDFs~\cite{Ji:2013dva,Ji:2014gla,Ji:2020ect}, pseudo-PDFs~\cite{Radyushkin:2017cyf,Orginos:2017kos,Radyushkin:2019mye}, lattice cross sections~\cite{Ma:2014jla,Ma:2017pxb}) provide the possibility of studying light-cone parton distributions directly with lattice simulations.
In particular, there 
were
 attempts of accessing the leading twist $B$-meson LCDA within 
 the
 quasi-distribution amplitude (quasi-DA) approach, either in coordinate space ~\cite{Kawamura:2018gqz} or momentum space~\cite{Wang:2019msf}. 
 Although the matching relation that links quasi-DA and LCDA
has  been 
 investigated, it is still not clear 
  how 
  one can 
   approach the continuum limit because of the existence of ultraviolet (UV) singularities. 

In this paper, we propose to deal with  the UV    singularities using 
the  pseudo-PDF approach~\cite{Radyushkin:2017cyf}.   Its essential  idea is that,  
 if the 
 operator is multiplicatively renormalizable, one can
 choose a proper ratio  that defines an  UV finite reduced Ioffe-time distribution.
To this end, we  will study 
the short-distance behavior of the
$B$-meson Ioffe-time distribution amplitude (ITDA) and construct a reduced ITDA. 

Using the results of the   one-loop calculation, we will show that the reduced-ITDA can be factorized into the
position-space LCDA and a hard function.  Furthermore, the UV finiteness allows the reduced ITDA calculated on the lattice to approach its continuum limit. 
This result  is crucial for building a practical method  of accessing $B$-meson LCDA on the lattice.

\section{$B$-meson Ioffe-time distribution amplitude}
Our starting object is 
a nonlocal heavy-light operator $O^{\mu}(z,0;v)\equiv\bar q(z)S(z,0)\gamma^{\mu}\gamma_5 h_v(0)$ in HQET, where $S(z,0)\equiv P \exp[i g z_{\nu}\int_0^1  d t A^{\nu} (t z)]$ is a Wilson line, and $h_v$ is the heavy quark field in HQET, with $v$ denoting its velocity, $v^2=1$
and 
 $h_v$ 
 satisfying 
 $\slashed v h_v=h_v$.
 The light quark is located at $z$, where $z$ is a spacelike vector.
By Lorentz covariance, the meson-to-vacuum matrix element can be parametrized as
\begin{align}
&\left\langle 0\left|\bar q(z)S(z,0) \gamma^{\mu}\gamma_5  h_v(0)\right| \overline B( v)\right\rangle\nonumber\\
=&i F(\mu)\left[v^{\mu} M_{B,v}(\nu, -z^2,\mu)+ z^{\mu} M_{B,z}(\nu, -z^2,\mu)\right], \label{eq:def}
\end{align}
where 
$M_{B,v}(\nu,\mu)$ and $M_{B,z}(\nu,\mu)$ are two scalar functions and $\nu\equiv v\cdot z$
will be referred to  
as the ``Ioffe-time'' of the $B$-meson 
(note that 
in the QCD case the Ioffe-time is the  inner product of momentum $p$ and $z$~\cite{Ioffe:1969kf,Braun:1994jq}). 
$F(\mu)$ is the decay constant of $B$-meson defined by the matrix element of the local current
\begin{align}
&\left\langle 0\left|\bar q(0) \gamma^{\mu}\gamma_5  h_v(0)\right| \overline B( v)\right\rangle=i v^{\mu} F(\mu).
\end{align}
Unlike the QCD case, decay constant in HQET is scale dependent.

When $z^2\to 0$, $M_{B,v}$ term gives the twist-2 distribution while $M_{B,z}$ is a higher-twist contribution. Note that the local limit has been included in the decay constant, so $z^2\to 0$ infers the light-cone limit for the distributions $M_{B,v}$ and $M_{B,z}$. 
Because  we are only interested in the leading-twist distribution at present, we 
rename $M_{B,v}$ as $M_{B}$ for short, and call $M_B(\nu,-z^2,\mu)$ the ITDA of the $B$-meson.

If $z$ is a lightlike vector, e.g.,  only the minus component of $z$ is nonzero, then  ITDA 
 will reduce to the light-cone ITDA  $\mathcal{I}_B^+(\nu,\mu)$, i.e., $M_B(\nu,0,\mu)=\mathcal{I}_B^+(\nu,\mu)$, which is actually the LCDA in position space.
The $B$-meson LCDA that appears in the factorization theorems of $B$-meson exclusive decay is defined by the Fourier transform of $\mathcal{I}_B^+(\nu,\mu)$~\cite{Grozin:1996pq}
\begin{align}
\phi^{+}_{B}(\omega, \mu)=\frac{v^+}{2 \pi} \int_{-\infty}^{\infty} d z^- e^{-i \omega v^+ z^-} {\mathcal{I}_B^+}(v^+ z^-, \mu).
\end{align}

There are no light-cone separations  on the Euclidean lattice, 
but 
as proposed in Refs. ~\cite{Braun:2007wv,Ji:2013dva}, 
one can study equal-time separations
\mbox{$z=(0,0,0,z_3)$.} 
The same idea may  also be applied  for the  $B$-meson LCDA. In this case, $\nu=-v_3 z_3$ and $z^2=-z_3^2$.
 One can choose the Lorentz index $\mu=0$ in Eq.~\eqref{eq:def},  so that the higher-twist part $z^{\mu} M_{B,z}$ disappears. 
 In the quasi-PDF-based  approaches ~\cite{Kawamura:2018gqz,Wang:2019msf} one deals with  
the $B$-meson quasi-DA $\tilde{\phi}_B^+(\omega,v_3,\mu)$  that can be expressed in terms of ITDA as
\begin{align}
\widetilde{\phi}_B^+ (\omega, v_3, \mu)=\frac{|v_3|}{2\pi}\int_{-\infty}^{\infty} dz_3 e^{i \omega v_3 z_3} M_B(-v_3 z_3,z_3^2,\mu).
\end{align}
A matching relation linking the quasi-DA and LCDA was derived in Ref.~\cite{Wang:2019msf}.

However, integration over the parameter $z_3$ present in both arguments
of the  ITDA $M_B(-v_3 z_3,z_3^2,\mu)$ mixes two distinct phenomena: 
the $\nu$-dependence  that governs the $\omega$-shape of the LCDA, and the  $z_3^2$-dependence 
that corresponds to the probing scale  for the LCDA.  
For this reason, we propose to proceed along the lines 
of the pseudo-PDF approach~\cite{Radyushkin:2017cyf,Radyushkin:2019mye} 
in which these phenomena are clearly  separated.

\section{Hard correction at one-loop}

Formally, 
the LCDA in coordinate space $\mathcal{I}_B^+(\nu,\mu)$ can be approached by taking $z^2\to 0$ limit of ITDA $M_B(\nu,-z^2,\mu)$. However, logarithmic dependence on $z^2$ will be generated when hard corrections of ITDA are included.
As a result,
the $z^2\to 0$ limit cannot be approached directly, and a perturbative matching is needed.

\begin{figure}
	\centering
	\includegraphics[width=1\linewidth]{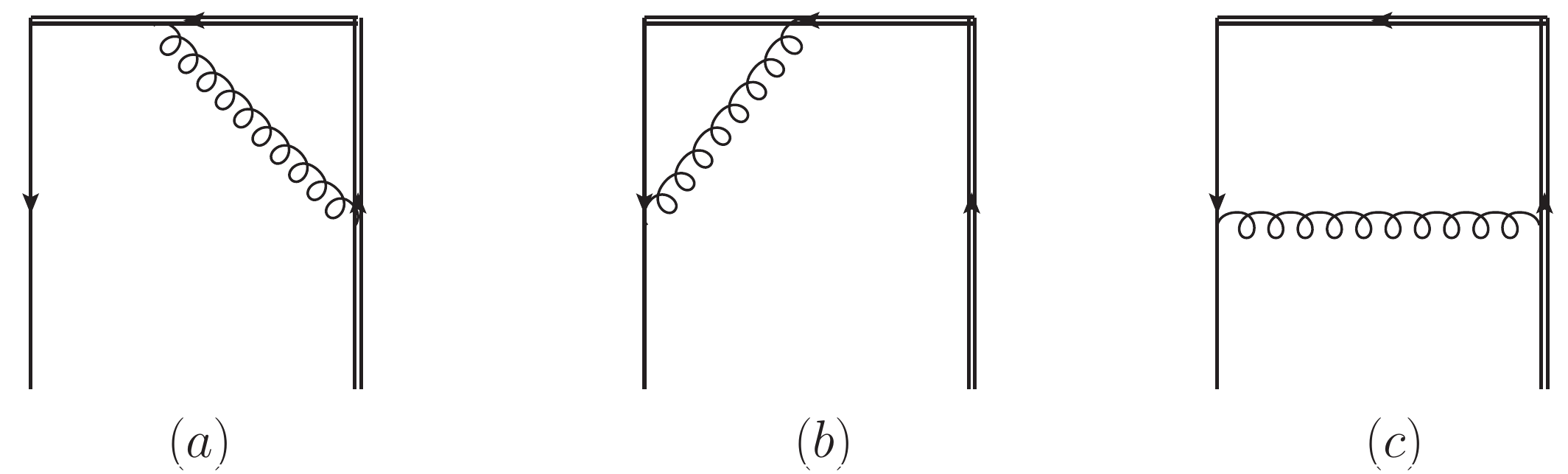}
	\caption{One-loop hard contribution to nonlocal heavy-light operator in HQET. The horizontal  double line represents the gauge link, while the vertical double line denotes the heavy-quark in HQET.}
	\label{fig:blcda}
\end{figure}

Under quantum correction, the hard part will be generated by gluon exchanges. As indicated in Refs.~\cite{Radyushkin:2017cyf,Radyushkin:2017lvu}, the hard contribution can be determined at operator level with coordinate representation. The Feynman diagrams are presented by Fig.~\ref{fig:blcda}.
A calculation has been performed in Ref.~\cite{Kawamura:2018gqz}, where the UV and IR singularities are regularized by dimensional regularization (DR). To distinguish the UV and IR singularities for ITDA, we will adopt Polyakov regularization~\cite{Polyakov:1980ca} for UV singularities, in which the gluon propagator in coordinate representation is replaced by $-g_{\mu\nu}/4\pi^2 z^2 \to  -g_{\mu\nu}/4\pi^2 (z^2-a^2)$. 
 Collinear singularities  are  regularized by 
 the mass of light quark
 $m$;
 the soft singularity is regularized by DR. We will work in Feynman gauge but the results are gauge invariant.

According to Eq.~\eqref{eq:def}, to study the hard contribution of $M_B(\nu,-z^2,\mu)$, one should consider the one-loop correction to both the nonlocal operator and decay constant. We consider the decay constant first.  Note that the $B$-meson decay constant in HQET is UV divergent and scale dependent, which is different from the pion decay constant case. Under Polyakov regularization, the one-loop hard correction to decay constant is
\begin{align}
&F(a)=F(a)^{(0)}\bigg[1-\frac{\alpha_s C_F}{2\pi}\bigg(\frac34\ln\frac{a^2 m^2 e^{2\gamma_E}}{4}+\frac{21}{8} \bigg)\bigg],\label{eq:dc:py}
\end{align}
where $\gamma_E$ is the Euler-Mascheroni constant and $F(a)^{(0)}$ denotes the decay constant without the hard correction, while $F(a)$ is the decay constant in which the hard correction is included.

Now we turn to the one-loop hard contribution of the nonlocal operator. To begin with, we consider the heavy quark and light quark self energies. Up to one-loop, we have
\begin{align}
 \delta Z_h=-\frac{\alpha_s C_F}{2\pi}\bigg(&\frac{1}{\epsilon_{\mathrm{IR}}}+\ln \frac{a^2 e^{2\gamma_E}}{4}+\ln4\pi\mu_{\mathrm{IR}}^2e^{-\gamma_E}\bigg),\nonumber\\
\delta Z_2=-\frac{\alpha_s C_F}{2\pi}    \bigg(&-\frac12\ln \frac{a^2 m^2 e^{2\gamma_E}}{4}+\frac{1}{\epsilon_{\mathrm{IR}}}\nonumber\\
&+ \ln\frac{4\pi\mu_{\mathrm{IR}}^2 e^{-\gamma_E}}{m^2} +\frac94\bigg)
\end{align}
for heavy and light quarks, respectively. Here 
$d=4-2\epsilon_{\mathrm{IR}}$ is the dimension of space-time in DR, and 
$\mu_{\mathrm{IR}}$ denotes the infrared (IR) scale associated with the soft singularity $1/\epsilon_{\mathrm{IR}}$. 
The self-energy of the gauge link has already been calculated in PDF case. The result reads~\cite{Chen:2016fxx,Radyushkin:2017lvu}
\begin{align}
\Gamma_{\Sigma}(z,a)= &\frac{\alpha_s C_F}{2\pi} \left(-\frac{\pi }{a}\sqrt{-z^2}+\ln\frac{-z^2}{a^2}+2\right)+\mathcal{O}(z^2). 
\end{align}

The heavy-quark Wilson line vertex is presented in Fig.~\ref{fig:blcda}(a). In HQET, the heavy quark can be expressed as a Wilson line along the $v$-direction.
At one-loop level, the exchange of gluon between Wilson lines along $v$ and $n$ directions contributes
\begin{align}
O^{\mu}&(z,0;v)
= \frac{\alpha_s  C_F}{2\pi} \bar q(z)\gamma^{\mu}\gamma_5 h_v(0)\nonumber\\
&\times \bigg[\ln a^2(\ln 2i v\cdot z-\frac12 \ln(-z^2))\nonumber\\
& +\frac14 \ln^2 (-z^2)
-\ln^2 2i v\cdot z-\frac{\pi^2}{6}\bigg]+\mathcal{O}(z^2) .
\end{align}
Note that the exchange of the gluon between the two Wilson lines generates a cusp singularity~\cite{Polyakov:1980ca}, which is represented  by $\ln a^2$. Another interesting 
feature 
here is that, because of the existence of cusp singularity,
 there is a double logarithmic dependence on $z^2$, 
 which is 
 very 
 different from the nucleon Ioffe-time distribution function
 case.  

The light-quark Wilson line vertex is presented in Fig.~\ref{fig:blcda}(b). This contribution is 
the same as  the vertex contribution in the PDF operator (see, e.g., Ref.~\cite{Radyushkin:2017lvu}). Direct calculation gives
\begin{align}
O^{\mu}(z,& 0;v)=\frac{\alpha_s C_F}{2\pi} \bigg\{\frac12(\ln\frac{-z^2}{a^2}-1) \bar q(z) \gamma^{\mu}\gamma_5 h_v(0)\nonumber\\
&-\int_0^{1} du  \bigg[ \ln\frac{-z^2 m^2 e^{2\gamma_E}}{4 }\frac{\bar u}{u}+\frac{\bar u+(2-u)\ln u^2}{u}\bigg]_+  \nonumber\\
&\times\bar q(\bar u z)   \gamma^{\mu}\gamma_5 h_v(0) \bigg\}+\mathcal{O}(z^2),
\end{align}
where $\bar u\equiv 1-u$. The plus distribution $[f(u)]_+$ is defined by $\int_0^1 du [f(u)]_+ T(u)\equiv \int_0^1 du f(u) [T(u)-T(u_0)]$, where $u_0$ is the pole of $f(u)$, and $T(u)$ is a test function.

Fig.~\ref{fig:blcda}(c) represents the contribution from interaction between light and heavy quarks. 
To calculate this contribution, we adopt the nonrelativistic approximation, 
where
the momentum of light quark is 
given by 
$p=m v$. Under this approximation, we have
\begin{align}
O^{\mu}&(z,0;v) =- \frac{\alpha_s C_F }{2\pi}  \bigg\{\bigg[\frac{2}{u}+\ln (i u  m v\cdot z e^{\gamma_E})\bigg]_+ \nonumber\\
&-\bigg[\frac{1}{\epsilon_{\mathrm{IR}}}-1+\ln \frac{4\pi\mu_{\mathrm{IR}}^2 e^{-\gamma_E}}{m^2}-\ln(i m v\cdot z e^{\gamma_E}  )\bigg]\delta(u)\bigg\}\nonumber\\
&\times \bar q(\bar u z)\gamma^{\mu}\gamma_5 h_v(0)+\mathcal{O}(z^2).
\end{align}
The Lorentz structure of the type $\gamma^{\mu}\slashed z\gamma_5$ has been neglected because it yields a higher-twist contribution to the ITDA.  One may notice that the box diagram has no $\ln z^2$ dependence,  so that it 
gives
the same 
contribution 
to the light-cone ITDA. 
This means
that the box diagram 
does not contribute to the matching relation.
This has been confirmed by the calculation in momentum space~\cite{Wang:2019msf}.

Adding all contributions together,
the soft IR singularities $1/\epsilon_{\mathrm{IR}}$, as well as the logarithmic dependence on the soft scale $\mu_{\mathrm{IR}}$, are canceled.
According to  Eqs.~\eqref{eq:def}  and \eqref{eq:dc:py}, one can derive the one-loop hard contribution of the ITDA $M_B(\nu,-z^2,\mu)$:
\begin{align}
M_B&(\nu, -z^2, a)=M_B(\nu)^{(0)}\nonumber\\
&+\frac{\alpha_s  C_F}{2\pi}  \bigg\{\bigg[ -\frac{\pi }{a}\sqrt{-z^2}+\frac32\ln\frac{-z^2}{a^2}+2\nonumber\\
&+ \ln a^2(\ln 2i \nu-\frac12 \ln(-z^2)) -\frac{\pi^2}{6}-\ln^2 2i \nu\nonumber\\
&+\frac14 \ln^2 (-z^2)  +\frac12\ln\frac{a^2}{4}  -\ln i \nu\bigg]M_B(\nu)^{(0)}\nonumber\\
&- \int_0^{1} dw \bigg[\frac{w}{\bar w} \ln\frac{-z^2 m^2 e^{2\gamma_E}}{4 }+\ln (i \bar w  m \nu e^{\gamma_E} )+\frac{2}{\bar w}\nonumber\\
&+\frac{w+(2-\bar w)\ln {\bar w}^2}{\bar w}\bigg]_+  M_B(w\nu)^{(0)} \bigg\}+\mathcal{O}(z^2) \label{eq:ITDA}.
\end{align}

\section{Reduced Ioffe-time distribution amplitude}
The hard contributions above involve  UV singularities that are 
regularized by $a$. In lattice computations, the matrix elements are calculated on discrete space-time. The UV singularities correspond to the singularities in the continuum limit (i.e., the lattice spacing $a\to 0$). Although the ITDA can be computed on the lattice, however, the UV divergences obstruct to approach the result in continuum space-time from lattice data. Thus one should renormalize the UV singularities for a practical lattice evaluation.

Based on the auxiliary field formalism~\cite{Gervais:1979fv}, it has been shown that the off-light-cone operator defining the $B$-meson quasi-DA is multiplicatively renormalizable~\cite{Wang:2019msf}.  So, the bare and renormalized operators are related by
\begin{align}
[\bar q(z)\gamma^{\mu}\gamma_5 h_v(0)]^{R}=Z(z\cdot v, z^2; \Lambda) \bar q(z)\gamma^{\mu}\gamma_5 h_v(0),
\end{align}
where $Z$ is a renormalization factor and $\Lambda$ denotes a cutoff. A similar equation can be written down for the decay constant. The operator with a superscript ``$R$'' denotes the renormalized operator while the operator without it denotes a bare one. The multiplicative renormalizability verified in Ref.~\cite{Wang:2019msf} will be the foundation of establishing a practically calculable quantity on the lattice.

Since the renormalization relation holds at operator level, it 
is valid 
for any matrix element of the operator. For example, one can replace $B$-meson state with the leading Fock state of $B$-meson. Similar to the definition of $B$-meson ITDA, such matrix element can be parametrized as 
\begin{align}
\langle 0|\bar q(z)\gamma^{\mu}\gamma_5 h_v(0) |b(v)\bar q(\omega v)\rangle =i v^{\mu}f(\mu) m_B(\omega \nu,-z^2),
\end{align}
where $f(\mu)$ and $m_B(\omega \nu, z^2)$ are the ``decay constant'' and ITDA of the Fock state $|b(v)\bar q(\omega v)\rangle$, respectively; $\omega v$ is the momentum carried by the light quark. Note that the higher-twist contribution that is  proportional to $z^{\alpha}$ has been neglected. $f(\mu)$ is defined through matrix element of local operator
\begin{align}
\langle 0|\bar q(0)\gamma^{\mu}\gamma_5 h_v(0) |b(v)\bar q(\omega v)\rangle =i v^{\mu}f(\mu).
\end{align}

As discussed above, the UV divergence only depends on the operator,
i.e., 
 the matrix elements of hadron state and its Fock state should involve the same UV structure. 
 Furthermore, the HQET operator is multiplicatively renormalizable, so the ratios of hadron and Fock-state matrix elements should be UV finite:
\begin{align}
\frac{F^R(\mu)}{F(\mu; a)}&=\frac{f^R(\mu)}{f(\mu; a)},\\
\frac{ F^R(\mu) M_B^R( \nu,-z^2)}{ F(\mu; a) M_B( \nu, -z^2; a)}&=\frac{f^R(\mu) m_B^R(\omega \nu, -z^2)}{f(\mu; a) m_B(\omega \nu, -z^2; a)}.
\end{align}
These relations indicate that for the ratio of meson and Fock-state ITDAs, the continuum limit  exists on the lattice, therefore the ratio can be evaluated with lattice simulations. For the sake of simplicity, 
we define a reduced ITDA $\overline{M}(\nu, -z^2)$ by dividing Fock-state ITDA at $\omega=0$:
\begin{align}
\overline{M}_B(\nu, -z^2)=\frac{M_B(\nu,-z^2;a)}{m_B(\omega \nu, -z^2;a)}\bigg\vert_{\omega=0}.\label{eq:def:reducedITDA}
\end{align}

Because Eq.~\eqref{eq:ITDA} is a general relation which is valid for ITDAs of both meson state and its leading Fock state ITDAs, 
one can immediately get the one-loop correction to the denominator of Eq.~\eqref{eq:def:reducedITDA}. The result reads
	\begin{align}
		m_B&(\omega \nu, -z^2, a)|_{\omega=0}=m_B(\omega \nu)^{(0)}|_{\omega=0}\nonumber\\
		&+\frac{\alpha_s  C_F}{2\pi} \bigg[ -\frac{\pi }{a}\sqrt{-z^2}+\frac32\ln\frac{-z^2}{a^2}+2\nonumber\\
		&+ \ln a^2(\ln 2i \nu-\frac12 \ln(-z^2)) -\frac{\pi^2}{6}-\ln^2 2i \nu\nonumber\\
		&+\frac14 \ln^2 (-z^2)  +\frac12\ln\frac{a^2}{4}  -\ln i \nu\bigg]m_B(\omega\nu)^{(0)}\bigg|_{\omega=0} \nonumber\\
		&+\mathcal{O}(z^2) \label{eq:denominator}.
	\end{align}	
Then, the one-loop correction of the reduced ITDA is
\begin{align}
\overline {M}_B&(\nu, -z^2)=\overline{M}_B(\nu)^{(0)} \nonumber\\
&- \frac{\alpha_s C_F }{2\pi}  \int_0^{1} dw \bigg[ \frac{w}{\bar w}\ln\frac{-z^2 m^2  e^{2\gamma_E}}{4 }+\ln (i \bar w  m \nu e^{\gamma_E} )\nonumber\\
&+\frac{2+w+(2-\bar w)\ln {\bar w}^2}{\bar w}\bigg]_+  \overline{M}_B(w\nu)^{(0)}+\mathcal{O}(z^2) \label{eq:RITDA}.
\end{align}	
It is easy to see 
that the linear and logarithm divergences that are related to the link are canceled; the cusp singularity that arises from the gluon exchange between heavy quark and link is canceled as well. 
This indicates the UV finiteness of the reduced ITDA. Thus it can be evaluated on the lattice,
 and the continuum limit can be approached.  Furthermore, it was pointed out that the denominator of the reduced-ITDA is not IR sensitive~\cite{Radyushkin:2017cyf}, thus the IR structure is not modified in the ratio. This is also verified by the one-loop result \eqref{eq:denominator}. 
 
In addition, the one-loop correction of the reduced ITDA is a plus-distribution. If we define a quasi-DA from
 the  reduced ITDA by taking Fourier transform with $z_3$, this will lead to a 
 rapidly 
 decreasing 
 behavior of the corresponding quasi-DA $\widetilde{\phi}(\omega,v_3,\mu)$ at large $\omega$.

On the lattice,  the denominator in the reduced ITDA 
 will  
be evaluated 
nonperturbatively. However, at short distances,  it can be calculated in perturbation theory. 
Thus,  this ratio defines a nonperturbative renormalization scheme for the $B$-meson ITDA.

\section{Matching relation}
The reduced ITDA and the $\overline{\mathrm{MS}}$ LCDA can be linked by a matching relation. Similar to the PDF case, one can use the nonlocal light-cone operator product expansion. 
To determine the hard function, we also need the one-loop correction to the light-cone ITDA, which can be extracted from the one-loop correction 
to  the light-cone operator~\cite{Zhao:2019elu}. The result reads
\begin{align}
\mathcal{I}_B^+(\nu,\mu)=&\mathcal{I}_B^+(\nu,\mu)^{(0)}\bigg\{1-\frac{\alpha_s C_F}{2\pi}\bigg[\ln^2(i {\mu} \nu e^{\gamma_E})\nonumber\\
&+\ln(i \mu \nu e^{\gamma_E})+\frac{5\pi^2}{24} \bigg] \bigg\}\nonumber\\
&+\frac{\alpha_s C_F}{2\pi}\int_0^1 dw \bigg[\frac{w}{\bar w}\ln\frac{\mu^2}{{\bar w}^2 m^2}-\frac{2}{\bar w}\nonumber\\
&-\ln(i\bar w e^{\gamma_E}m \nu)\bigg]_+  \mathcal{I}_B^+(w\nu,\mu)^{(0)}+\mathcal{O}(\alpha_s^2). 
\end{align}
One can find that the singularities regularized by $\ln m^2$ are 
the same for the 
reduced-ITDA and light-cone ITDA. Then, a matching formula
for reduced ITDA and light-cone ITDA can be written down:
\begin{align}
&\overline{M}_B(\nu,z_3^2)
=\mathcal{I}_B^+(\nu,\mu)\nonumber\\ &+\frac{\alpha_s C_F}{2\pi}\bigg\{\bigg[\ln^2(i\tilde\mu \nu ) 
+\ln(i\tilde\mu \nu  )+\frac{5\pi^2}{24}\bigg] \mathcal{I}_B^+(\nu,\mu)\nonumber\\
&- \int_0^1 du\bigg[\frac{u}{\bar u}\bigg(\ln\frac{ z_3^2 \tilde\mu^2 }{4}+1\bigg)+2\frac{\ln\bar u}{\bar u}\bigg]_+ \bigg\}\mathcal{I}_B^+(u\nu,\mu)\nonumber\\
&+\mathcal{O}(\alpha_s^2),
\end{align}
where $\tilde\mu\equiv \mu e^{\gamma_E}$. We have chosen $z=(0,0,0,z_3)$ so that the reduced ITDA can be computed on the lattice.
This relation allows one 
to convert
the reduced ITDA 
calculated on the lattice to the LCDA in coordinate representation. 
The
Fourier transformation 
of the latter
enters the factorization theorems of $B$-meson exclusive decay.

Finally,  
let us take a look at the evolution equations 
for 
the ITDAs. Since the light-cone ITDA does not depend on $z^2$, one can write down the $z^2$-evolution equation for the reduced ITDA:
\begin{align}
\frac{d}{d\ln z_3^2}\overline{M}_B(\nu,z_3^2)=-\frac{\alpha_s C_F}{2\pi}\int_0^1 du \bigg[\frac{u}{\bar u}\bigg]_+ \overline M_B(u\nu,z_3^2).
\end{align}
On the other hand, the reduced ITDA does not depend on $\mu$.
Thus, 
by taking the derivative with respect to $\ln \mu$ on both sides, one can get the renormalization group equation for light-cone ITDA:
\begin{align}
\mu\frac{d}{d\mu}\mathcal{I}_B^+(\nu,\mu)=&-\frac{\alpha_s C_F}{\pi}\bigg\{\bigg[\ln(i \tilde{\mu}\nu)+\frac12\bigg]\mathcal{I}_B^+(\nu,\mu)\nonumber\\
&-\int_0^1 du\bigg[\frac{u}{\bar u}\bigg]_+\mathcal{I}_B^+(u\nu,\mu)\bigg\}.
\end{align}
By including the anomalous dimension of the decay constant, one will find that the above equation reproduces the RGE for heavy-light light-cone operator (see, e.g., Refs.~\cite{Kawamura:2010tj,Zhao:2019elu}).

\section{Implementations on the lattice}
In recent lattice calculations, the typical lattice spacing $a$ is around $0.1$~fm. The Compton wavelength of 
the $b$-quark is  much smaller than the lattice spacing, 
$m_b\gg 1/a$. Hence HQET is a natural framework to study $B$-meson on the lattice.

The renormalization of full HQET is complicated, even in continuum theory. In fact, the full HQET Lagrangian is not renormalizable. 
However, since 
the operator we are going to measure is 
taken in the
$m_b\to \infty$
limit, 
we can restrict ourselves 
to 
 the static approximation of HQET.  At the lowest order of $1/m_b$ expansion, there will be no higher dimensional operators getting mixed in,
  and the renormalization property is simple. The higher dimensional operators in lattice theory can also been excluded under static approximation, hence the continuum limit of the reduced ITDA can be approached, without considering the operator mixing. 

Similar to the lattice calculation of hadron Ioffe-time distribution functions~\cite{Orginos:2017kos}, a possible way to get the ITDA is to calculate $M_B(-v_3 z_3, z_3^2)$ for several values of $v_3$, and then to fit the data by a function of $\nu$ and $z_3^2$. A proper framework might be the leading-order moving HQET~\cite{Mandula:1991ds}. The meson and Fock state decay constants should also be calculated on the lattice, or using the phenomenological result. 
The Fock state ITDA at $\omega=0$,
 i.e.,  $m_B(0\cdot \nu, z_3^2)$ 
 necessary for the construction of reduced ITDA and   
 should be calculated on the lattice as well.

We note that, 
 in practical lattice HQET, 
  a result with large noise-to-signal ratio might be expected.
  Still,  
  a rough evaluation of the $B$-meson LCDA with lattice methods is 
  of great value because there is very little knowledge on the $B$-meson LCDA, even from first principle calculations.

\section{Summary}

 To access 
$B$-meson leading-twist light-cone distribution amplitude from lattice QCD computations, 
we have proposed the  approach based on the strategy of reduced Ioffe-time distributions. 
The reduced Ioffe-time distribution amplitude of a $B$-meson is constructed by the ratio of meson ITDA and the ITDA of meson's leading Fock state, in which the light-quark momentum is zero. According to the multiplicative renormalizability of the off-light-cone operator, the ratio is UV finite; hence, 
one can approach its continuum limit from the lattice data. A matching relation that maps LCDA to the reduced ITDA is also derived. These results provide a basis 
for a practical computation of $B$-meson LCDA with lattice methods.

\section*{Acknowledgments}
This work is supported by Jefferson Science Associates, LLC under  U.S. DOE Contract \#DE-AC05-06OR23177 and by U.S. DOE Grant \#DE-FG02-97ER41028.

\end{document}